\documentclass[11pt,preprint]{aastex}
\usepackage{psfig}
\received{}
\revised{}
\accepted{}
 
\journalid{}{}
\articleid{}{}
\paperid{}
\shortauthors{Perrett et al.}
\shorttitle{Substructure in the M31 GCS}

\newcommand{\kms}{\hbox{km$\,$s$^{-1}$}}
\newcommand{\invkms}{\hbox{km$^{-1}\,$s}}

\begin{document}

\title{Substructure in the Andromeda Galaxy Globular Cluster System}

\author{K. M. Perrett, D. A. Stiff, D. A. Hanes,}
\affil{Department of Physics, Queen's University, Kingston, ON K7L 3N6, Canada}
\and\author{T. J. Bridges}
\affil{Anglo-Australian Observatory, Epping, NSW, 1710 Australia}

\begin{abstract}

In the most prominent current scenario of galaxy formation, galaxies
form hierarchically through the merger of smaller systems.  Such
mergers could leave behind dynamical signatures which may linger long
after the event.  In particular, the globular cluster system (GCS) of
a merging satellite galaxy may remain as a distinct sub-population
within the GCS of a massive galaxy.  Using the latest available
globular cluster velocities and metallicities, we present the results
of a search for grouping in the GCS of our nearest large spiral galaxy
neighbor, M31.  A modified friends-of-friends algorithm is used to
identify a number of possible merger remnants in projected position,
radial velocity and [Fe/H] parameter space.  Numerical simulations are
used to check that such merger remnants are indeed plausible over the
timescales of interest.  The identification of stellar streams
associated with these groups is required in order to confirm that they
represent merger remnants.

\end{abstract}


\section{Introduction}

In hierarchical structure formation scenarios, galaxies form at least
in part from the merger and accretion of smaller clumps of matter
\citep[e.g.,][]{sea78, col00}.  Clumping in the early universe is
believed to have resulted from small perturbations in the primordial
density field, instabilities that triggered the growth of structure
from the ``bottom up''.  Such primordial clumps would then evolve into
dwarf galaxies like those observed in the present epoch, or may merge
to form more massive galaxies. 

There are indications that such mergers are ongoing, as is evident
from observations of the Sagittarius dwarf galaxy's current
interaction with the Milky Way \citep{iba95, doh01, new02}.  There are
also indications of tidal distortions in M31's dwarf satellite NGC~205
\citep{wal87, ben91}.  Furthermore, a wide-field imaging survey of M31
by \citet{iba01} has revealed the presence of a giant stream of
metal-rich stars within the halo of M31. This feature may have
originated from tidal interactions with M31's dwarf companions,
NGC~205 and M32.  Either the process of satellite accretion is not
uncommon, or we are observing our Galaxy and M31 at particularly
unique stages of their evolution.  Recent studies by \citet{rei02} and
\citet{fer02} have also found signs of substructure and tidal trails
in the M31 halo which support the possibility of past accretion
events.

Globular clusters (GCs) have been identified within some of the dwarf
satellites of the Milky Way \citep{dac95,hod99,oh00} and M31
\citep{gre00}, as well as in the dwarf members of distant galaxy
clusters \citep[e.g.][]{dur96, mil98}.  Based on their positions,
distances and radial velocities, there are four Milky Way globular
clusters which are believed to be associated with the Sagittarius
dwarf \citep{iba94, iba97, bel03}.  Ultimately, these clusters ---
M54, Arp 2, Terzan 7 and Terzan 8 --- will have their orbits
randomized by the strong potential of the Milky Way and they will
presumably take their place as {\it bona fide} members of the Galactic
globular cluster system.

Therefore, in the context of hierarchical galaxy formation, one would
expect that the globular cluster systems (GCSs) of massive galaxies
could include the cannibalized globular clusters of past-accreted
dwarfs. For instance, this seems to be a plausible mechanism for
building up the GCSs of giant elliptical galaxies
\citep[e.g.][]{cot98}.  Such a scenario raises the question of whether
the globular cluster systems of merged dwarfs would remain distinct
dynamical entities or instead become hopelessly mixed with the larger
galaxy's GCS. If they remain recognizable over significant
time-scales, satellite GCS remnants could provide us with a powerful
dynamical probe of the formation and evolution of galaxies.


The first significant effort at distinguishing sub-groups within the
GCS of an external galaxy was performed for the M31 system by
\citet{ash93}.  These authors suggested that an apparent
sub-clustering in the positions and velocities of M31 globular
clusters represents a surviving signature of progenitor gas clouds
from which the galactic halo itself may have formed.  Ashman \& Bird
employed a technique in which globular clusters were grouped based on
deviations from global mean velocities and the velocity dispersions
between each cluster and its $N$ nearest neighbors.  Their sample
included 144 M31 GCs with spectroscopic data from \citet{huc91}, many
of which had large velocity uncertainties ($\gtrsim\pm 50$ \kms).

In many respects, the M31 globular cluster system provides an ideal
target in the search for substructure and grouping.  More than 435
confirmed globular clusters have been identified in M31 \citep{bar00}.
A recent spectroscopic study of the M31 GCS by \citet{per02} has now
pushed the number of clusters with radial velocities and metallicities
beyond 300.  In addition, \citet{per02} quote typical velocity
uncertainties for the majority of the clusters of $\pm
12$~\kms, a significant improvement over many previous studies.  The
time is right to take a closer look for signatures of hierarchical
formation within the M31 globular cluster system.

There are several important things to consider in this type of study.
One must define what is to be regarded a significant ``grouping'' in
the available parameter space.  In other words, appropriate grouping
criteria must be established {\it a priori}.  Furthermore, one must
investigate the likelihood that any observed groups may be mere
coincidental associations rather than actual substructure.  A greater
difficulty is the interpretation of the significance of the grouping
results: do the groups represent true merger remnants that have
survived to the present epoch?

In this paper, we develop a technique to identify potential dynamical
remnants within a galaxy's globular cluster system.  This
group-finding algorithm is described in
Section~\ref{sec:groupfinding}.  In Section~\ref{sec:m31groups}, we
apply the group-finding technique to the sample of M31 GCS positions,
velocities and metallicities from \citet{per02}.  We identify several
potential groups of clusters in this sample.  Monte Carlo simulations
are performed on randomized globular cluster populations to
characterize the significance of the identified groups.
Section~\ref{sec:simulations} describes our use of N-body simulations
to investigate whether these groups could represent actual merger
remnants.  Finally, the results of this study are summarized in
Section \ref{sec:conclusions}.


\section{The Technique of Group Identification}
\label{sec:groupfinding}

In most applications, simulated data has the advantage over
observational data in that it provides full three-dimensional
positional and dynamical information.  Observational data is, of
course, less ideal: we must cope with projected positions and
line-of-sight velocities and we can, at best, make generalized
assumptions about the true dynamical behavior of any system.
Additionally, every observed parameter carries with it an uncertainty,
the value for which may vary from object to object even within the
same dataset.  These are some of the considerations that must be made
when devising a group-finding algorithm for observational data.

\subsection{Group-Finding Algorithm}

The basic technique used in our group identification is a
friends-of-friends algorithm.  This algorithm links together particles
that fall within a specified distance in parameter space.  Each
distinct set of joined particles constitutes a group.  In the case of
the M31 globular cluster system observations, the parameter space is
defined by $X$ and $Y$ (projected) positions, radial velocities $v_r$,
and metallicity [Fe/H].

The appropriate weighting of these parameters must be determined in
advance in order to set the group-linking criteria.  Since each
parameter has units, each must be converted to an equivalent unitless
quantity on the same scale before calculating distances in parameter
space.  For example, if $X$ and $Y$ each span a range of $100$~arcmin,
the velocities range over 600~\kms, and metallicity spans only 3~dex,
an appropriate metric must be devised such that the group is not
defined principally based on [Fe/H] owing to its comparatively small
numerical range.  In essence, the metric puts all of the parameters on
to the same effective scale.  In addition to the metric, the
group-finding algorithm requires a linking length which defines the
maximum allowable distance between successive group members in the
specified parameter space.

Therefore, for each GC, the algorithm links a nearby GC to it if
\begin{equation}
(\Delta s)^2 \equiv \sum_i \gamma_i^2(\Delta x_i)^2 < {\cal L}^2,
\end{equation}
where $\Delta s$ is the total separation (in parameter space) of the
neighbor to the GC in question, $\gamma_i$ is the scaling or
weighting factor defined for parameter $i$, $\Delta x_i$ is the
separation between the GC and its neighbor in parameter $i$, and
${\cal L}$ is the linking length.  Traditionally, ${\cal L}$ is
specified as a fraction, $\Lambda$, of the mean inter-particle
spacing for the entire system:
\begin{equation}
{\cal L} = \Lambda \overline{\Delta s}
\end{equation}
The advantage of using a friends-of-friends technique over average
density methods is that it can follow elongated structures, features
that may be expected in tidal disruptions.

The error bars associated with each input parameter are accommodated
by the addition of an ``uncertainty ellipsoid'' surrounding each GC in
the available parameter space.  A GC's parameter-space neighbor is
linked with it if the minimum distance between the uncertainty
ellipsoids is less than the specified linking length.


\subsection{Metric and Linking-Length Determinations}
\label{sec:metric}

In order to define a suitable metric to be used in the definition of a
group, a method was devised to test the group-finding algorithm on
simulated associations using a range of scalings and linking lengths.
Groups of objects with various population sizes, spatial distributions
(e.g. spherical, linear and curved streams), galactocentric positions
and orientations were created with general parameters similar to those
of Local Group dwarf galaxies.  The simulated groups were given
internal velocity dispersions of $\sigma_v=6$ to $12$~\kms, mean
metallicities of $\langle$[Fe/H]$\rangle=-1.2$ to $-1.6$ dex, and
[Fe/H] spreads of $\sigma_{\rm [Fe/H]}=0.3$ to $0.5$ dex
\citep{mat98,dac00}.

The GCs in these simulated groups were then superimposed over a
background designed to mimic a realistic distribution of M31 globular
clusters.  This background of several hundred GCs consisted of the
following components:
\begin{itemize}
\item[1.] a random, metal-poor, spherical halo distribution, 
\item[2.] a rotating, metal-rich thick disk distribution inclined at
$12.3^\circ$ to the line-of-sight, and
\item[3.] a rotating, intermediate-metallicity ellipsoidal bulge
distribution.
\end{itemize}

Projected positions, radial velocities and metallicities for the
fabricated group members and additional background objects were
analysed using the group-finding algorithm.  The procedure was cycled
through a range of linking lengths and parameter scalings to determine
the optimal combination for group retrieval.  A minimum group
membership of $n_{GC}=4$ clusters was specified to define a group.

Setting the $X$ and $Y$ scaling to \mbox{$\gamma_X=\gamma_Y=1.0$}
arcmin$^{-1}$ to define the baseline, it was found that the best
recovery rate occurred when \mbox{$\Lambda=0.1$} and the velocities
were scaled by \mbox{$\gamma_v=0.13$} \invkms\ for cases when
metallicities were not considered in the group recovery.  With the
addition of metallicity information, the optimal velocity and [Fe/H]
scalings were found to be \mbox{$\gamma_v=0.14$ \invkms} and
\mbox{$\gamma_{\rm [Fe/H]}=9.7$ dex$^{-1}$}, respectively, with a
linking length given by \mbox{$\Lambda=0.09$}.  These optimal linking
length and parameter scaling values were used to search for grouping
in the globular cluster system of M31.

\section{Substructure in the M31 GCS}
\label{sec:m31groups}

\subsection{The GCS Sample}

The best available spectroscopic sample of $\gtrsim 300$ GC positions,
radial velocities and metallicities as defined in \citet{per02} was
examined for signs of grouping.  Quoted errors on individual
velocities and metallicities were used to define the uncertainty
ellipsoids about each data point in parameter space.  Uncertainties on
the target positions were typically around 0.2 -- 1.0 arcseconds;
these were deemed to be negligible and thus position uncertainties
were not incorporated in the analysis.  

The group-finding algorithm was used to analyze the data based on the
parameter scalings and linking lengths determined in
Section~\ref{sec:metric}.  Again, a minimum of four clusters was
required in order to define a group.  A search for groups was
conducted in position/velocity/metallicity parameter space for the
sample of 301 globular clusters with published spectroscopic
metallicities.  Roughly one dozen groups were initially flagged within
the GC population.  The robustness of these groups was then tested
against small ($\sim 10\%$) changes in the specified parameter
scalings and linking length.  Groups that were not found to be stable
against these perturbations were rejected.

At this point it should be noted that the inclusion of metallicity
information may not necessarily improve the effectiveness of the
group-finding results. The [Fe/H] uncertainties for many of the M31
globular clusters remain quite large ($\pm 0.3 - 0.9$ dex).  This
reduces the significance of metallicity as a linking parameter due to
the increased limits of each uncertainty ellipsoid in this dimension.
Furthermore, it has been demonstrated that the dwarf galaxies of the
Local Group exhibit a wide variety of star formation histories and
metallicities \citep[see][and references therein]{mat98}.  For
example, \citet{gre99} report a large \mbox{($\sim 1$ dex)} spread in
stellar metallicities within the M31 satellites And~VI and And~VII.
Similar spreads have been found in the globular cluster systems of
some dwarf elliptical galaxies (J. Lotz, private communication).  As a
result, proximity in metallicity is not necessarily a very suitable
criterion for group identification within a hierarchical formation
scenario.

For these reasons, the complete available sample of 321 M31 globular
cluster positions and velocities was also examined for grouping
without the inclusion of [Fe/H] information.  Again, roughly a dozen
potential groups were identified and then tested for stability against
small changes in the grouping criteria.  The majority of the resulting
groups correspond to those found when metallicity was included as a
linking parameter.

\citet{per02} identified significant bimodality in their sample of M31
GC metallicities, in confirmation of earlier reports made by
\citet{huc91} and \citet{ash93}.  The centrally concentrated, rapidly
rotating, metal-rich cluster system is consistent with a bulge
population \citep{per02}.  The group-finding results in the full GCS
data sample both with and without metallicities revealed a couple of
relatively large groups that correspond to a sizable fraction of this
metal-rich bulge population.  A third grouping analysis was therefore
also performed on the sample of 231 GCs in the metal-poor population
as defined by \citet{per02}.  Metallicity was not used as a linking
parameter in this test.  Several tentative groups persisted within the
central region even within this metal-poor population, yet the
memberships of all but one of these groups were not robust to small
variations in the linking criteria.  Ultimately, groups with average
galactocentric radii within the central $10\arcmin$ of the galaxy were
rejected due to the higher probability of false detections, an effect
which is confirmed by the Monte-Carlo simulations presented later in
Section~\ref{sec:montecarlo}.  Additionally, any cannibalized dwarf
GCSs would be extremely difficult to recover within the dense inner
regions of the galaxy; this effect will also be noted in the numerical
simulation results discussed in Section~\ref{sec:simulations}.

\subsection{Grouping Results}

\begin{figure}
\plotone{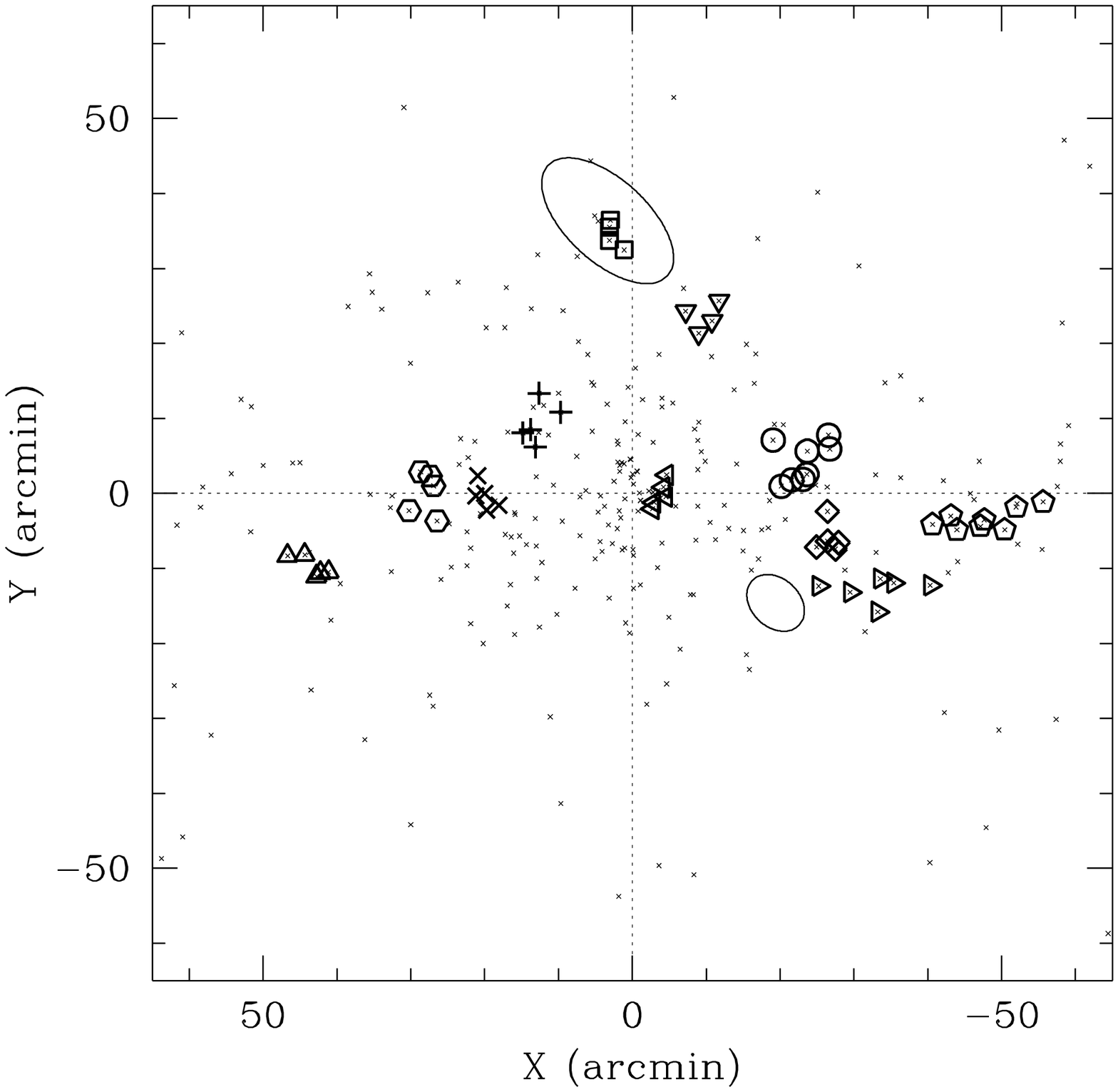}
\caption{The positions of the groups identified in the M31
globular cluster system.  Globular cluster group members are shown
with matching symbols and their data are presented in
Table~\protect\ref{tab:groups}.  M31's dwarf companions (NGC~205 and
M32) are shown by the large ellipses.}
\label{fig:groups}
\end{figure}

In the M31 GCS, there is evidence for the presence of ten unique
globular cluster groups of $n_{GC}\geq 4$ members with mean
galactocentric radii beyond $10\arcmin$.  The positions of these
groups are shown in Figure~\ref{fig:groups}, and the data for the
group members are provided in Table~\ref{tab:groups}.  Five of these
groups were identified in all three search cases described above.
Three additional groups (Groups 7, 8 and 10) resulted from a search of
the full sample of 321 globular clusters without the inclusion of
[Fe/H] information.  Two further groups can be considered as tentative
identifications: Group 3 was found only when metallicity was included as
a linking parameter, and Group 4 resulted only from the search of the
metal-poor population.  All of the groups listed in
Table~\ref{tab:groups} are stable to small perturbations in the
linking criteria.

An eleventh group is also listed in Table~\ref{tab:groups};
despite the fact that this group lies close to the galaxy center, the
mean radial velocity of its group members ($-155$~\kms) is
considerably higher than the systemic velocity of M31 ($-300$~\kms).
This group deviates significantly from the GCS rotation curve
determined by \citet{per02}, and in fact is moving counter to the
general rotation at that major axis position.

The association of a group with NGC~205 is clearly evident in
Figure~\ref{fig:groups} (Group 5 of Table~\ref{tab:groups}).  The mean
velocity of the NGC~205 group is \mbox{$v_r = -259$~\kms} with
$\sigma_v = 78$~\kms, a result that is consistent with NGC~205's
systemic velocity of $-242\pm 3$~\kms\ \citep{RC3}.

It is interesting to note that none of the groups recovered in this
analysis overlap with the associations found by \citet{ash93}.  This
may be partly attributable to the larger sample of higher precision
velocities used here.  \citet{ash93} also indicated that their method
may produce false hits in the presence of GCS rotation. In an effort
to reduce this affect, they applied a KMM mixture-modelling technique
to remove the rotating metal-rich (disk/bulge) population from their
grouping analysis, thus decreasing their sample from 144 GCs to 95
``halo'' objects \citep{huc91}.  However, \citet{per02} found that
even the metal-poor cluster population demonstrates significant
rotation, a factor which may also have biased the \citet{ash93}
grouping results.

The majority of the clusters in the inner and outer subsystems found
by \citet{sai00} are included in the metal-rich (bulge) population
inferred from the KMM results of \citep{per02}.  The bulk of inner
subsystem found by \citet{sai00} was recovered here as two central
groups; the memberships of these inner groups were not stable and
these are not listed in Table~\ref{tab:groups} for reasons discussed
previously.

\begin{deluxetable}{clrrrc}
\tabletypesize{\tiny}
\tablewidth{0pt}
\tablecaption{M31 globular cluster groups\label{tab:groups}}
\tablehead{
 \colhead{ID} & \colhead{GC} & \colhead{$X$ ($\arcmin$)}  
 & \colhead{$Y$ ($\arcmin$)} & \colhead{$v_r$ (\kms)}  & \colhead{[Fe/H]}
}
\startdata
1
 & B484-S310 & 46.69 & -8.31  & $-104 \pm 12$ & $-1.95\pm 0.59$\\
 &      B480 & 44.36 & -8.18  & $-135 \pm 12$ & $-1.86\pm 0.66$\\  
 &     DAO84 & 42.76 & -11.08 & $-192 \pm 12$ & $-1.79\pm 0.72$\\ 
 & B376-S309 & 42.22 & -10.64 & $-142 \pm 20$ & $-2.18\pm 0.99$\\  
 & B374-S306 & 41.14 & -10.50 & $ -96 \pm 12$ & $-1.90\pm 0.67$\\
\hline
2
 & B229-S282 & 30.25 & -2.30 & $-103 \pm 88$ & $-1.81 \pm 0.74$  \\
 &     DAO66 & 28.68 &  2.88 & $-148 \pm 12$ & $-1.82 \pm 0.26$  \\
 &     DAO65 & 27.37 &  2.32 & $-130 \pm 12$ & $-1.80 \pm 0.36$  \\
 & B216-S267 & 26.90 &  1.02 & $ -84 \pm 12$ & $-1.87 \pm 0.39$  \\
 & B223-S278 & 26.45 & -3.68 & $-101 \pm 12$ & $-1.13 \pm 0.51$  \\
\hline
3\tablenotemark{\dagger} 
  & B203-S252 & 21.19 & -0.36 & $-199 \pm 12$ & $ -0.90 \pm 0.32$  \\
  & B190-S241 & 20.95 &  2.38 & $ -86 \pm 12$ & $ -1.03 \pm 0.09$  \\
  & B198-S249 & 19.98 &  0.02 & $-105 \pm 12$ & $ -1.13 \pm 0.30$  \\
  & B206-S257 & 19.75 & -2.23 & $-151 \pm 23$ & $ -1.45 \pm 0.10$  \\
  &      B200 & 18.07 & -1.57 & $-153 \pm 12$ & $ -0.91 \pm 0.61$  \\
\hline
4\tablenotemark{\dagger} 
 & B141-S197 & 14.83 &  8.06 & $-180 \pm 12$ & $-1.59 \pm 0.21$ \\
 & B137-S195 & 13.79 &  8.50 & $-215 \pm 12$ & $-1.21 \pm 0.29$ \\
 &     DAO58 & 13.13 &  6.19 & $-125 \pm 12$ & $-0.87 \pm 0.07$ \\
 &      B102 & 12.63 & 13.34 & $-236 \pm 12$ & $-1.57 \pm 0.10$ \\
 & B105-S166 & ~9.71 & 10.83 & $-238 \pm 12$ & $-1.13 \pm 0.32$ \\
\hline
5
 &  B11-S63 & ~3.12 &  33.74 & $-157 \pm 52$ & $-1.54\pm 0.34$\\
 & B328-S54 & ~3.11 &  35.50 & $-243 \pm 23$ & $-1.51\pm 0.28$\\
 & B324-S51 & ~2.95 &  36.42 & $-299 \pm 35$ & $-1.39\pm 0.40$\\
 &   B9-S61 & ~1.11 &  32.47 & $-335 \pm 52$ & $-1.57\pm 0.26$\\
\hline
6\tablenotemark{\ast} 
  & B126-S184 & -2.76 & -2.04 & $-182 \pm 14$ & $-1.20 \pm 0.47$ \\
  &      NB91 & -2.93 & -1.19 & $-187 \pm 10$ & $-0.71 \pm 0.33$ \\ 
  &      NB83 & -4.23 &  0.84 & $-150 \pm 14$ & $-1.26 \pm 0.16$ \\ 
  & B106-S168 & -4.54 & -0.43 & $ -90 \pm 43$ & $-0.86 \pm 0.68$ \\ 
  &  B86-S148 & -4.64 &  2.48 & $-168 \pm 21$ & $-1.74 \pm 0.17$ \\ 
\hline
7
 &  B13-S65 & ~-7.21 & 24.32 & $-409 \pm 12$ & $-1.01 \pm 0.49$  \\ 
 &  B16-S66 & ~-8.99 & 21.35 & $-406 \pm 12$ & $-0.78 \pm 0.19$  \\ 
 &  B12-S64 & -10.79 & 22.98 & $-358 \pm 12$ & $-1.65 \pm 0.19$  \\
 &   B4-S50 & -11.72 & 25.70 & $-373 \pm 12$ & $-1.26 \pm 0.59$  \\
\hline
8
 &      V31 & -19.06 & 7.11 & $-433 \pm 12$ & $-1.59 \pm 0.06$  \\
 &     V216 & -20.17 & 0.96 & $-465 \pm 12$ & $-1.15 \pm 0.26$  \\
 &  B33-S95 & -21.57 & 1.78 & $-439 \pm 12$ & $-1.33 \pm 0.24$  \\
 &  B31-S92 & -23.12 & 1.88 & $-400 \pm 12$ & $-1.22 \pm 0.40$  \\
 &  B28-S88 & -23.64 & 2.54 & $-434 \pm 12$ & $-1.87 \pm 0.29$  \\
 &     B453 & -23.69 & 5.64 & $-446 \pm 12$ & $-2.09 \pm 0.53$  \\
 &      B15 & -26.57 & 7.78 & $-460 \pm 12$ & $-0.35 \pm 0.96$  \\
 &    DAO39 & -26.73 & 5.92 & $-478 \pm 12$ & $-1.22 \pm 0.41$  \\
\hline
9
 & B57-S118 & -24.94 &  -7.16 & $-437 \pm 12$ & $-2.12\pm 0.32$\\
 &  B34-S96 & -26.43 &  -2.40 & $-540 \pm 12$ & $-1.01\pm 0.22$\\
 &     B458 & -26.44 &  -6.37 & $-521 \pm 12$ & $-1.18\pm 0.67$\\
 & B49-S112 & -27.49 &  -7.41 & $-481 \pm 12$ & $-2.14\pm 0.55$\\
 &    DAO48 & -27.91 &  -6.55 & $-490 \pm 12$ & $-2.01\pm 0.99$\\
\hline
10
 & B81-S142 & -25.22 & -12.36 & $-430 \pm 12$ & $-1.74 \pm 0.40$  \\
 & B66-S128 & -29.50 & -13.18 & $-389 \pm 12$ & $-2.10 \pm 0.35$  \\
 & B65-S126 & -33.24 & -15.81 & $-378 \pm 12$ & $-1.56 \pm 0.03$  \\
 & B43-S106 & -33.58 & -11.38 & $-414 \pm 12$ & $-2.42 \pm 0.51$  \\
 & B40-S102 & -35.38 & -11.94 & $-463 \pm 12$ & $-0.98 \pm 0.48$  \\
 & B342-S94 & -40.35 & -12.27 & $-479 \pm 12$ & $-1.62 \pm 0.02$  \\
\hline
11
 &  B18-S71 & -40.63 &  -4.15 & $-585 \pm 12$ & $-1.63\pm 0.77$\\
 &     B448 & -43.13 &  -3.01 & $-552 \pm 12$ & $-2.16\pm 0.19$\\
 &     B335 & -43.95 &  -4.87 & $-514 \pm 12$ & $-1.05\pm 0.26$\\
 &   BoD195 & -47.14 &  -4.38 & $-552 \pm 12$ & $-1.64\pm 0.19$\\
 & B327-S53 & -47.65 &  -3.51 & $-528 \pm 12$ & $-2.33\pm 0.49$\\
 &     B443 & -50.41 &  -4.85 & $-532 \pm 12$ & $-2.37\pm 0.46$\\
 & B319-S44 & -51.99 &  -1.81 & $-535 \pm 12$ & $-2.27\pm 0.47$\\
 & B315-S38 & -55.62 &  -1.13 & $-559 \pm 12$ & $-1.88\pm 0.52$\\
\hline
\enddata
\tablenotetext{\ast}{Central group}
\tablenotetext{\dagger}{Tentative group}
\end{deluxetable}

\subsection{Monte-Carlo Simulations}
\label{sec:montecarlo}

A series of Monte-Carlo simulations were performed to investigate the
likelihood that the identified M31 globular cluster groups are mere
chance associations on the sky.  Randomized populations of 321
globular clusters were generated based on the globular cluster
positions and radial velocities from \citet{per02}.  Metallicities
were not included in the randomized tests due to the large [Fe/H]
uncertainties in the data.  Furthermore, the eight non-tentative outer
groups in our sample were found without incorporating any metallicity
information in the group finding.

First, the cumulative distribution observed in projected position
along the major axis of the galaxy was used to select a random cluster
X position with roughly the same underlying distribution as that seen
in the M31 GCS.  Next, the cumulative distribution in projected minor
axis location of the 70 nearest objects in X was used to generate a
randomized (but realistic) value of Y position for each target.
Similarly, the cumulative distribution in radial velocity for the 70
nearest objects in X was used to generate a randomized $v_r$ for each
target.  This procedure generated a population of globular clusters
which shares similar observed properties with the observed data ---
including spatial distribution and rotation profile --- while
eliminating any small-scale correlations that may exist in the
dataset.

\begin{figure}
\plotone{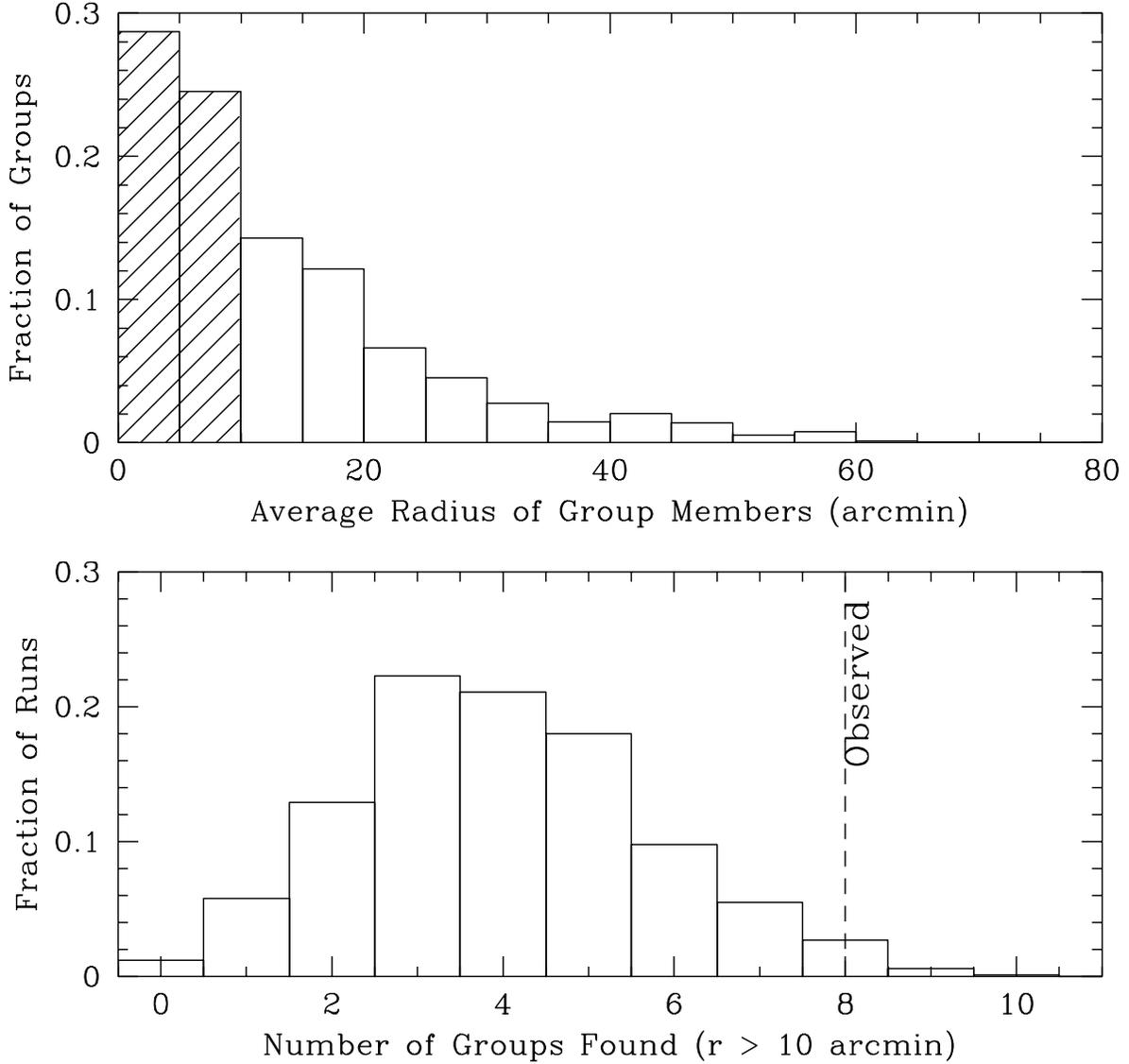}
\caption{The grouping results in 1000 randomized
populations. The top panel shows the distribution in mean
galactocentric radius of the false groups found in the randomized
data.  Roughly 50\% of the false hits originated within the central
$10\arcmin$ of the galaxy (shaded bins).  The lower panel shows the
number distribution of groups with galactocentric radii greater than
$10\arcmin$ that were identified in the randomized data.  The number
of groups found in the actual M31 GCS data is shown by the dashed
vertical line.}
\label{fig:h1000}
\end{figure}

One thousand realizations of the randomized GC populations were
performed and each was tested for grouping in position and velocity
parameter space.  The results of these Monte Carlo tests are shown in
Figure~\ref{fig:h1000}. The top panel of Figure~\ref{fig:h1000} shows
the radial distribution of the groups found in the randomized data.
Roughly half of these ``false hits'' were recovered within the central
$10\arcmin$ of the galaxy, a region which appears to be too dense to
yield reliable group detections.  The fact that the spatially compact
(metal-rich) GC population demonstrates rapid rotation also adds to
the likelihood of detecting chance conglomerations in position-velocity
parameter space.  Therefore, neglecting all of the groups with mean
radii within $10\arcmin$ of the galaxy center, we obtain the number
distribution of random-data groups shown in the lower panel of
Figure~\ref{fig:h1000}.

If we exclude the two tentative groups and the central group, the
remaining eight groups observed in the actual M31 dataset
(Table~\ref{tab:groups}) lie above $\sim97\%$ of the randomized runs.
This suggests that the substructure found in the M31 GCS is
significant overall. If we further exclude the group associated with
NGC 205, this result still lies above $\sim91\%$ of the randomized
runs.  Therefore, it is unlikely that {\em all} of the groups are
chance associations, although the majority may indeed be.  For
example, there is nearly a $60\%$ chance that three or more of the groups
are not real.  With the currently available data, it is not possible
to determine if the individual groups are dynamically bound
structures, nor can we gauge the specific degree of internal
contamination by background clusters.


\section{Modelling the Fate of the Globular Cluster Systems}
\label{sec:simulations}

Based on the discovery of apparent substructure in the M31 GCS, it is
worthwhile to investigate the plausibility that the proposed groups
are a result of merger events.  To accomplish this, it is instructive
to consider the fate of a dwarf satellite's small globular cluster
population during a merger with a massive galaxy.  N-body simulations
were performed to follow the evolution of a dwarf galaxy GCS over
timescales of $\tau \gtrsim 1$~Gyr as it interacts with an external
potential representing a massive spiral galaxy. The form of the
galactic potential and the modelling of the dwarfs are discussed in
Sections~\ref{sec:galpot} and \ref{sec:dwarf}, respectively.  The
interpretation of the simulation results is presented in
Section~\ref{sec:simresults}.

\subsection{The Galactic Potential}
\label{sec:galpot}

In this study, the galactic potential was assumed to be rigid and
time-independent, and was built from three components: a spheroidal
bulge potential \citep{her90}, a \citet{miy75} thick disk, and a
logarithmic halo.  The forms of these potentials were based on
the modelling of the Sagittarius dwarf by \citet{joh95} and
\citet{hel01}.  The potentials are:
\begin{eqnarray}
\phi_{\rm disk}  &=& \frac{-GM_{\rm disk}}{\sqrt{R^2+(a+\sqrt{z^2+b^2})^2}}\\
\phi_{\rm bulge} &=& \frac{-GM_{\rm bulge}}{(r+c)}\\
\phi_{\rm halo}  &=& v_{\rm halo}^2 \ln{(r^2+d^2)}
\end{eqnarray}
where $G$ is the gravitational constant, $r$ and $R$ represent radii
in spherical and cylindrical coordinates, respectively, and $z$ is the
height above the galactic plane.  The masses ($M$), scale lengths
($a$, $b$, $c$, $d$) and velocity ($v_{\rm halo}$) in the above
potentials were selected to reproduce the characteristics of the Milky
Way.  The components of our Galaxy are better understood than those of
M31 and we can expect the overall effects of satellite disruption
within M31 to be quite similar.  The adopted component parameters are
provided in Table~\ref{tab:MWpot} and have been selected following the
work of \citet{joh98}.

\begin{table}
\begin{center}
\caption{Galaxy model parameters}
\label{tab:MWpot}
\vspace{6pt}
\begin{tabular}{cc}\hline\hline
{\bf Parameter} & {\bf Adopted Value} \\ \hline
$M_{\rm bulge}$ & $3.4\times 10^{10} M_\odot$ \\
$M_{\rm disk}$ & $1\times 10^{11} M_\odot$ \\
$v_{\rm halo}$ & $128$~\kms \\
$a$ & $6.5$~kpc \\
$b$ & $0.26$~kpc \\
$c$ & $0.7$~kpc \\
$d$ & $12.0$~kpc \\ \hline
\end{tabular}
\end{center}
\end{table}

The representation of the massive galaxy by a fixed potential neglects
any exchange of energy between the satellite and its parent galaxy.
However, typical masses of dwarf satellites are several orders of
magnitude smaller than that of either the Milky Way or M31.  As a
result, the effects of dynamical friction and energy exchange between
the satellite and the massive galaxy can be neglected over the
timescales of interest here \citep{hel01, bin87}.

\subsection{The Model Dwarf}
\label{sec:dwarf}

The dwarf satellite galaxy in each simulation was modelled as a
Hernquist sphere with a density distribution given by
\begin{equation}
\rho(r)=\frac{M_{\rm sat}}{2\pi}\frac{s}{r}\frac{1}{(r+s)^3}
\end{equation}
\noindent \citep{her90}, which corresponds to a cumulative mass 
distribution of
\begin{equation}
\label{eq:hernmass}
M(r)=M_{\rm sat}\frac{r^2}{(r+s)^2} .
\end{equation}
$M_{\rm sat}$ is the total satellite mass and $s$ is a scale length
that defines the half-mass radius of the distribution:  setting
$M(r_{1/2}) = M_{\rm sat}/2$ in Equation~\ref{eq:hernmass} gives
$r_{1/2}=(1+\sqrt{2})s$.  This analytical mass model is an
approximation of the de Vaucouleurs $R^{1/4}$ profile applicable to
galaxies and bulges \citep{dev53}.

For each satellite, roughly $10\,000$ particles were distributed
according to this density profile.  Dwarf galaxies of several masses
were considered within the range from $10^7 M_\odot$ to $10^9
M_\odot$.  Ten of the particles were given a somewhat more extended
distribution about the dwarf and were designated to be its globular
clusters; these GCs were each assigned masses of $5\times10^5
M_\odot$.  The remaining mass of the simulated dwarfs was equally
divided among the rest of the particles.  The number of particles used
to represent the dwarf was sufficient to model its disruption, while
being small enough to efficiently run the required number of
simulations.

A series of $85$ simulations were run in order to explore a reasonable
range of parameter space by altering the initial conditions of the
incident satellite.  Table~\ref{tab:initcond} lists the ranges of
initial conditions that were tested.  The total dwarf masses ($M_{\rm
sat}$) were selected to span a range between a typical dwarf
spheroidal and a dwarf elliptical and include the mass of its 10
globular clusters. The range in satellite half-mass radii ($R_{1/2}$)
was chosen to correspond to typical values of dwarf galaxy core radii
--- a reasonable assumption if light follows mass.  $R_{\rm GC}$ is
the scale radius of the dwarf's globular cluster system.  The
satellites were assigned a starting galactocentric distance ($D$) and
a position angle ($\theta$) above the plane of the disk.  The
azimuthal angle was taken to be $\phi=0$, since the potential is
axisymmetric. The parameters $\alpha$ and $\beta$ dictate the initial
partition of velocity in the $XZ$ and $XY$ plane, respectively, in
units of circular velocity such that $\alpha^2 + \beta^2 = 1$ for
circular orbits.

These N-body simulations were performed using the Barnes-Hut
{treecode}\footnote{Treecode version 1.4 distributed by J. Barnes.} and
modified to include an external potential.  The {treecode} algorithm
is based on the commonly-used hierarchical force method described by
\citet{bar86}.  

\begin{table}
\begin{center}
\caption{Initial conditions for N-body simulations}
\label{tab:initcond}
\vspace{6pt}
\begin{tabular}{cc}\hline\hline
{\bf Parameter} & {\bf Initial Conditions}\\ \hline
$M_{\rm sat}$ & $10^7 - 10^9 M_\odot$\\
$R_{1/2}$   & $0.2 - 0.4$~kpc \\
$R_{\rm GC}$   & $ (1 - 2) R_{\rm 1/2}$ \\
$D$            & $20 - 100$~kpc \\
$\theta$       & $0 - 90^\circ$ \\
$\alpha$       & $0 - 0.9$ \\ 
$\beta$        & $0 - 0.9$ \\ \hline
\end{tabular}
\end{center}
\end{table}

It should be noted that the simulations presented herein are intended
to be illustrative, and {not} to reproduce the detailed
characteristics of this complex interaction.  The addition of evolving
potentials for the bulge, disk and halo components and the
incorporation of effects such as dynamical friction, disk-shocking,
and gas dynamics would permit a more comprehensive investigation of
the merger event itself.


\subsection{Grouping in the Simulation Results}
\label{sec:simresults}

Using the scaling and linking lengths determined in the previous step
($\Lambda=0.1$, $\gamma_v=0.13$ \invkms), the N-body simulation results were
analysed to investigate the likelihood of identifying the cannibalized
dwarf's globular clusters as a group in parameter space.  The
three-dimensional positions and velocities were projected into an
orientation similar to M31 ($i=77.7^\circ$), and were superimposed on
the background components described in Section~\ref{sec:metric}.  The
group-finding algorithm was then applied to the simulation results at
various timesteps.

Two parameters, $Q$ and $Q_{max}$, were defined in order to quantify
the retrieval efficiency of the satellite's GCS:
\begin{equation}
\label{eq:Q}
Q=\frac{1}{N}\sum_k{ \left(\frac{n_{GC}}{n_{g}}\right)_k}
\end{equation}
\begin{equation}
Q_{max} = {\rm max}\left(\frac{n_{GC}}{n_{g}}\right)_k
\end{equation}
where $N$ is the total number of groups containing at least one of the
dwarf's original globular clusters, ${n_{GC}}$ is the number of the
globular clusters in group $k$, and ${n_{g}}$ is the total number of
objects (including background) that were identified in group $k$.  The
sum in $Q$ is calculated only over those groups that contain one or
more globular clusters and reflects the general contamination of the
GC groups by background objects.  $Q_{max}$ is the maximum ratio of
the number of satellite GCs in a group to the total number of objects
allocated to that group, and therefore indicates whether a {\it bona
fide} group has been found.  For perfect group recovery with no
contamination, $Q=Q_{max}=1$.

To demonstrate, let us examine a particular simulation with an initial
dwarf position angle $90^\circ$ ($\alpha=0.9$, $\beta=0$) with $M_{\rm
sat} = 10^9 M_\odot$ at a timestep of 1.2 Gyr.  The
group-finding algorithm returned globular cluster group memberships as
follows: 6 of the 10 dSph GCs were in the same group and had 1
additional background object included, 1 GC was allocated to a
background group with 3 additional members, 1 GC was allocated to a
background group with 4 additional members, and 2 GCs were not
identified as members of any group.  For this case with $N=3$ groups
containing one or more globular clusters:
\begin{displaymath}
Q = \frac{1}{3}\,{\left[ \frac{6}{7} + \frac{1}{4} + \frac{1}{5}
\right]} = 0.44
\end{displaymath}
\begin{displaymath}
Q_{max} = {\rm max}\left(\frac{6}{7}, \frac{1}{4}, \frac{1}{5} \right)
= 0.86
\end{displaymath}
The high $Q_{max}$ indicates that it is very likely that we can detect
a group that contains a significant fraction of the disrupted
satellite's GCS, here with only $14\%$ contamination by background
objects.  The moderate to low $Q$ ratio reflects the fact that some of
the satellite's GCs were allocated to other groups to which they do
not truly belong.

The orbital periods of the satellite debris within the potential of
the parent galaxy are of order 1~Gyr.  The probability of detection
clearly depends to a large degree on the relic's location in its
orbit, as there is a decreased likelihood of finding a group if the
GCS debris is fore or aft of the bulge, or obscured in the rotating
disk particles.  Therefore, the likelihood that the detection of a
group in a galaxy's GCS reveals the remnant of a real accretion event
is comparatively low if the group is superimposed over the dense part
of the disk or the bulge of the galaxy.  Not surprisingly, satellite
GCS remnants are also most easily observable at large galactocentric
radii due to their significantly lower degree of disruption over the
relevant timescales.

\begin{figure}
\plotone{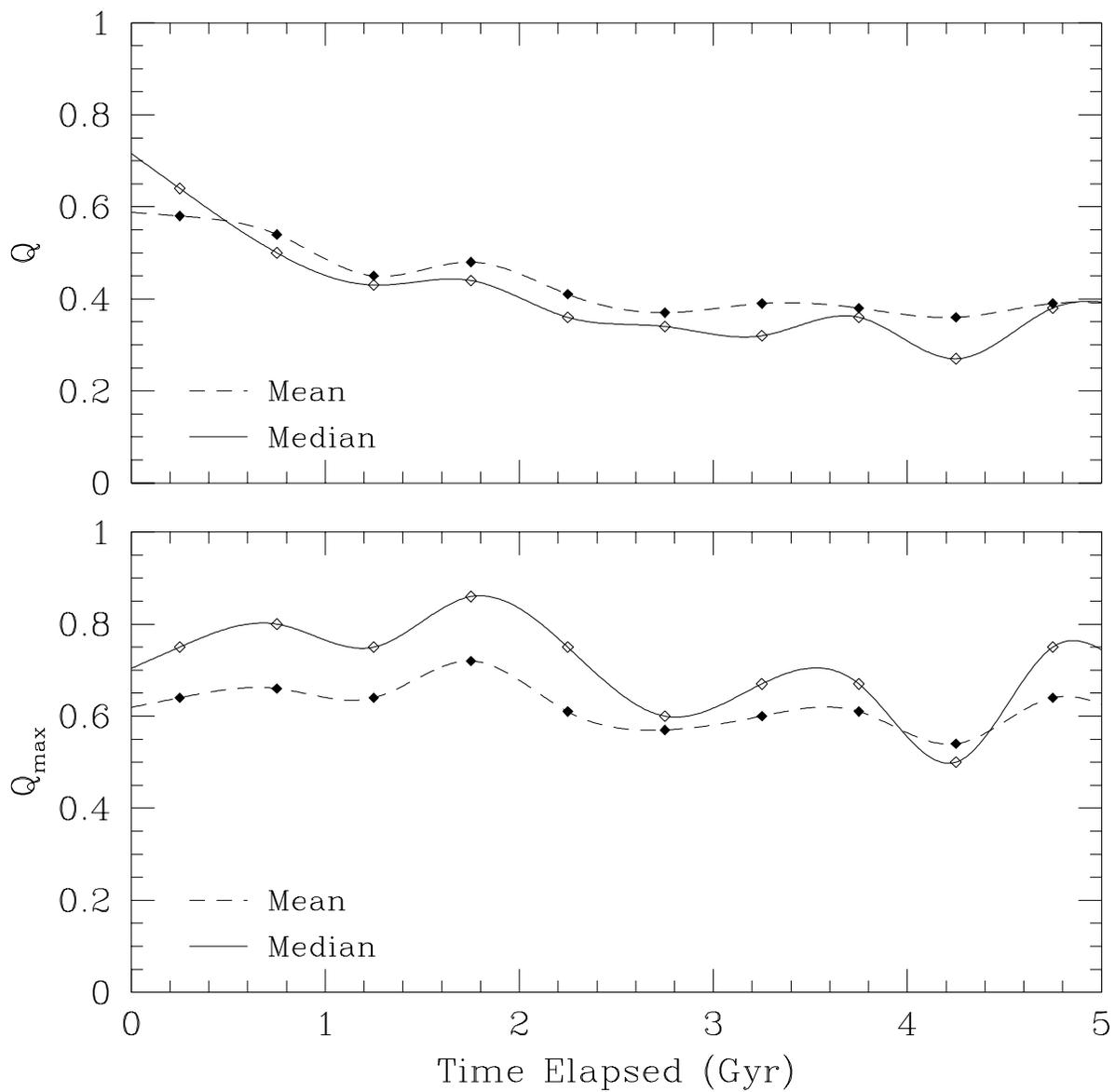}
\caption{The average $Q$ ratio (top plot) and $Q_{max}$
values (bottom plot) of a series of simulations with a range of
initial conditions at an initial distance of $40$~kpc.  The dashed and
solid lines show smoothed fits to the mean and median values in
timestep bins, respectively.} 
\label{fig:SimOut}
\end{figure}

In reality, we rarely know the orbital parameters of past (or even
present) satellite galaxies.  We can therefore examine the average $Q$
values of a large sample of the simulations having a variety of
initial position angles and incident velocities to get some idea of
the likelihood of group detection for a satellite on a random orbit.
The mean and median $Q$ and $Q_{max}$ values for satellites with an
initial galactocentric radius of $40$~kpc are shown in
Figure~\ref{fig:SimOut}.  The average and median $Q$ values decrease
with time, as expected.  The $Q$ results indicate that it is not
likely that we can recover the complete GCS of the satellite dwarf in
a single uncontaminated group.  However, the $Q_{max}$ plot shows that
one significant grouping of clusters associated with the cannibalized
dwarf can generally be found up to several Gyr after the initial
encounter.  


\section{Conclusions}
\label{sec:conclusions}

The numerical simulation results presented in this study reveal that
it is possible to identify some fraction of a disrupted satellite's
initial GCS based on grouping in projected position and radial
velocity.  This can be accomplished for timescales of
$\tau\,\gtrsim\,1$~Gyr after the initial encounter, although generally
not without some significant degree of background contamination in the
group membership.

In the Milky Way, there are three-dimensional data available that
trace a distinct stellar stream associated with the Sagittarius dwarf
merger.  It is then relatively easy to pick out any globular clusters
that coincide with this stream in both distance and velocity, although
even here the results are somewhat controversial \citep[for example,
see][]{min96}.  At the distance of M31, we have projected data and are
only now beginning to uncover signs of stellar streams and
substructure \citep{iba01}.  These observed streams extend well beyond
the spatial limits of the current cluster sample (on a line connecting
NGC~205 and M32), and thus cannot be directly matched with any of the
GC groups identified in this study.  Large-area surveys
\citep[e.g.,][]{mas02, lee02} will help to provide the necessary
additional stellar astrometry and to reveal any new GC candidates in
M31.  Nonetheless, \citet{lyn95} demonstrated that it is possible to
use streams of globular clusters to trace the orbits of potential
merger remnants within the Milky Way.  Streams or groups of globular
clusters in nearby massive galaxies may also provide a viable means of
identifying the orbits of any recently merged satellites.  

We have found significant signs of grouping in projected position and
radial velocity within the globular cluster system of M31.  Do these
M31 globular cluster groups represent the signatures of past accretion
events?  Without accurate distances or dynamics, the correlation of
M31 GCS groups with past accretion events remains tenuous.  In followup
studies, the identification of a stellar stream coincident with a
grouping of globular clusters would lend the necessary credibility to
the notion that a galaxy's GCS --- and by extension the galaxy itself
--- formed at least partly in a hierarchical manner.


\section{Acknowledgments}

The authors wish to thank Larry Widrow and Geraint Lewis for useful
comments and related discussions.  The authors are also grateful to
Josh Barnes for making his treecode software available, and to the
anonymous referee for the helpful suggestions.  This work was
supported in part by funding from the Natural Sciences and Engineering
Research Council of Canada.



\end{document}